# A Generic Minimal Discrete Model for Toroidal Moments and Its Experimental Realization

Hong Xiang[1,*], Lixin Ge[1,*], Liang Liu[1,*], Tianshu Jiang[2],

Z. Q. Zhang[2], C. T. Chan[2,†], and Dezhuan Han[1,†]

[1]Department of Applied Physics, Chongqing University, Chongqing 401331, China

[2] Department of Physics and Institute for Advanced Study, The Hong Kong University of Science and Technology, Clear Water Bay, Kowloon, Hong Kong, China

[*] These authors contributed equally to this work.

[†] Correspondence address: dzhan@cqu.edu.cn or phchan@ust.hk

It is well known that a closed loop of magnetic dipoles can give rise to the rather elusive toroidal moment. However, artificial structures required to generate the necessary magnetic moments are typically optically large, complex to make and easily compromised by the kinetic inductance at high frequencies. Instead of using magnetic dipoles, we propose a minimal model based on just three aligned discrete electric dipoles in which the occurrence of resonant toroidal modes is guaranteed by symmetry. The advantage of this model is its simplicity and the same model supports toroidal moments from the microwave regime up to optical frequencies as exemplified by a three-antenna array and a system consisting of three nano-sized plasmonic particles. Both the microwave and high-frequency configurations exhibit non-radiating "anapoles". Experiments in the microwave regime confirm the theoretical predictions.

Toroidal moments, which can appear in different disciplines of physics[1-5], serve as a third family of multipole moments in addition to conventional electric and magnetic multipole moments. A typical model of a toroidal dipole is a torus on which the currents circulate along the meridians, or a folded solenoid. Although the radiation patterns of toroidal multipoles are indistinguishable from the conventional electric or magnetic multipoles[6,7], their characteristics such as the resonance frequencies and quality factors can be much different from those of conventional multipoles[8,9]. Furthermore, microstructures exhibiting resonant toroidal modes can be employed as the building blocks of metamaterials[10-13], leading to possible applications in nano-antenna design[14], high-Q cavity sensing[15], and photon emission engineering[16].

While toroidal moments have been discussed in the literature for many years, it is only very recently that metamaterials exhibiting such moments have been successfully made and measured. Most of the previous schemes achieve toroidal moments by introducing an array of magnetic dipoles aligned along a circle, where the magnetic dipoles originate from the split-ring resonators[8,17-20] or other artificial structures[21-29]. However, it is well known that the magnetic resonance undergoes a saturation effect at the optical frequencies for the split-ring resonator[30,31]. Therefore, achieving optical toroidal response using a system of subwavelength scale resonators with magnetic resonances is a challenge. Absorption is another issue, which is particularly problematic for toroidal moments which are typically very weak and the resonance peaks are easily smeared out by absorption effects. One possible way to avoid these problems is to employ high-index dielectric metamaterials[32]. Since the refractive indices of the normal dielectrics are relatively low at the optical frequencies, the toroidal modes may appear in the Mie scattering regime rather than in the subwavelength scale. It is also revealed that the toroidal moments play a considerable role in the electromagnetic (EM) scattering for the optically large dielectric/metallic particles[33]. The interplay of electric and toroidal dipole moments can even give rise to a non-radiating "anapole"[34]. As is known, higher multipole moments will become more significant as the size of the scatterer increases. The existence of resonant toroidal modes in the subwavelength scale is still of both theoretical and practical interest.

On the other hand, localized surface plasmons supported by metal particles are one of the most significant phenomena in plasmonics[35-37]. Metal particles hence can merge electronics and photonics at the nanoscale[38,39], and in particular, they can act as electric dipole resonators at the optical frequencies. Noting that magnetic dipoles are generally more difficult to realize than electric dipoles, we show here that toroidal modes can in fact emerge from just three *resonant electric dipoles* placed close to each other. This *minimal* model consists of nothing more than three electric dipoles aligned parallel to each other. The existence of a resonant toroidal dipole is guaranteed by the inversion symmetry inherent in the system. The advantage of this model is its simplicity and more importantly, it is a generic model in the sense that it gives rise to toroidal moments in different frequency regimes. We will give two examples: the dipole antennas in the microwave regime and metal nanoparticles at the optical frequencies. In these systems, toroidal modes can be excited by external waves, and the resonant behavior can be characterized by a dynamic eigenmode analysis[40-42]. In the system of dipole antennas, the characteristics of resonant toroidal modes can be observed from both the scattering spectra and the spatial field profiles in the near field. The experimental results agree excellently with the numerical simulations as well as the theoretical analysis. In the plasmonic system, the resonant toroidal modes can emerge from the tri-particle assembly independent of the geometrical details of the constituent particles as long as they exhibit dipolar resonances. The rather amazing phenomenon of non-radiating "anapoles" naturally arise from our minimal model as a result of the destructive interference between electric and toroidal dipoles. The simplicity of our model can take toroidal modes a step closer to potential applications.

## RESULTS

**Resonant toroidal mode realized by three electric dipoles.** Our idea is presented in Fig. 1. The blue loops with arrows in Fig. 1a show schematically the current pattern that can generate a toroidal dipole ***T*** indicated by the red arrow. The magnetic dipole moments generated by these current loops also form a loop, as shown by the green arrows. Most of the toroidal moments discussed in the literature were achieved through realizing the conceptual method of Fig. 1a using an array of artificial magnetic dipole resonators. Another way to generate toroidal moments is to align some discrete current elements (e.g. antennas or metal nanoparticles) from head to tail along this current flow, as represented by the yellow arrows in Fig. 1a. Physically, one just needs to put the electric dipoles on the sites of the yellow arrows. As we shall show below, this structure can be further reduced to a *minimal* version, namely, just three electric dipoles with inversion symmetry shown in Fig. 1b. Using the symmetry analysis, the transverse modes of this minimal model can be categorized into one anti-symmetric, and two symmetric modes, denoted respectively by $A$, $S_1$ and $S_2$ modes (see, e.g., the insets of Fig. 2a,c for the corresponding field patterns). The $A$ mode can be utilized to realize a resonant magnetic dipole mode and provide a way to achieve negative effective permeability at the optical frequencies[43]. The $S_1$ mode has all the three dipoles oscillating in phase. The $S_2$ mode, in which the side dipoles oscillate out of phase with the central one, is of particular interest in this study. As illustrated in Fig. 1b, the current associated with this mode can be viewed as a current source from the central dipole that is split into two parts and flows back through the side ones. If the outflowing current of the central dipole is equal to the inflowing currents of the side dipoles, it can be regarded as a pure toroidal dipole as inferred from the definition. In general, the instantaneous total current will not be zero since the sum of the central outflowing current and the side inflowing currents are not zero. However, they can almost cancel each other under some conditions as will be shown both numerically and analytically below.

**Toroidal mode in the system of three antennas.** In the first example, the half-wave dipole antennas are adopted to serve as the "current elements" of Fig. 1b. A linear response theory for the system of antennas is developed (see Methods for details). With no loss of generality, we choose the radius of the antennas as $r_0=0.1$ mm, the length of the side antennas as $L_1=L_3=15$ mm, and the length of the central antenna as $L_2=14$ mm. As sketched in the upper left inset in Fig. 2a, the three antennas point along the $z$ axis with their centers aligned on the $y$ axis. The spacing between the adjacent antennas is $d=6$ mm. Since this system does not have the cylindrical symmetry, the EM scattering is dependent on the incident angle $\theta$. We assume that the incident wave is a plane wave of the form $\boldsymbol{E}_{\text{in}} = \hat{e}_z E_0 e^{i(k_x x + k_y y)}$, which is polarized along the $z$ axis and propagating in the $x$-$y$ plane. The wire antennas can be treated as perfect electric conductors (PECs) in the microwave regime. Two typical configurations, with $\theta=\pi/2$ (glancing incidence, Fig. 2a,b), and $\theta=0$ (normal incidence, Fig. 2c,d) are considered.

The numerical results calculated using the COMSOL package for the scattering efficiency $Q_{sca}$ are shown in Fig. 2a for the glancing incidence. The scattering efficiency $Q_{sca}$ is defined by the scattering cross section divided by the geometric cross section. The geometric cross section of the three antennas is $2r_0L_1$ for the glancing incidence, and $2r_0(2L_1+L_2)$ for the normal incidence. There are two peaks marked by the blue dots: one is at 8.75 GHz with a relatively low quality factor and the other is at 9.48 GHz with a high quality factor. The corresponding field distributions of the normal component $E_x$ on the plane which has a vertical distance 3 mm to the antennas' plane, are plotted in the corresponding insets. By examining the field patterns, one can distinguish these two peaks. The one at 8.75 GHz is dominated by the anti-symmetric mode (*A* mode), while the one at 9.48 GHz corresponds to the $S_2$ mode showing the characteristics that the fields on the side antennas are out of phase with the central one. From the previous discussions on the nature of these modes, we expect that the $S_2$ mode at 9.48 GHz should possess a large toroidal moment, while the *A* mode at 8.75 GHz is a magnetic dipole mode. This can be further verified by the calculation of the radiation spectra for different multipoles. Using the surface currents from the simulation, the radiation spectra due to different multipole moments, including ***P***, ***T***, ***M***, ***Q***$_e$, ***Q***$_m$, are shown in Fig. 2b. These multipole moments are defined in the Methods. At around the frequency 9.48 GHz, the radiation of the toroidal dipole moment plays a dominant role, which demonstrates that a toroidal dipole resonance can indeed exist in this simplest discrete system. The radiation spectrum for the magnetic dipole moment has a peak at 8.75 GHz corresponding to the peak of the *A* mode in Fig. 2a, as expected.

For the normal incidence, the scattering efficiency and radiation spectra for different multipoles, are also shown in Fig. 2c,d for comparison. The peak at about 9.48 GHz ($S_2$ mode) can still be observed in Fig. 2c. However, the peak at 8.75 GHz (*A* mode) in Fig. 2a no longer shows up in Fig. 2c, as the anti-symmetric mode cannot be excited under the normal incidence. There is another peak with the lowest quality factor at 10.0 GHz. The spatial field profile shows that this is the ordinary electric dipole mode with all the three antennas oscillating in phase. This electric dipole mode does not appear in Fig. 2a under the glancing incidence since it cannot be excited resonantly as the incident fields on the three antennas are not synchronized in phase. The radiation spectra are shown in Fig. 2d, and not surprisingly, the electric dipole radiation spectrum has a broad peak centered at 10.0 GHz. However, the radiation from the electric dipole now exceeds that from the toroidal dipole at about 9.48 GHz ($S_2$ mode). This means that the net current of these three antennas (note that $\mathbf{P} \sim \int \mathbf{j} d^3 r$) are not negligible under the normal incidence and the radiation from the electric dipole becomes stronger than that from the toroidal dipole in the far field.

The simulated results are confirmed by experiments. In the experiments, three antennas (copper wires) with the radius $r_0=0.1$ mm are placed on a flat substrate of EPS foam (dielectric constant $\varepsilon \sim 1$ in the microwave region), as sketched in the upper inset of Fig. 3. The incident plane wave is generated by a horn antenna, and a monopole antenna is placed on top of the sample with an effective vertical distance 3 mm to detect the normal field component $E_x$. The scanning in the $y$-$z$ plane can be conducted with a minimum step of 0.08 mm. A vector network analyzer (Agilent 5232A) is used to feed the horn antenna and record the signals from the detector antenna. Absorbing materials are used to reduce the reflection from the environment.

The measured near field spectra of $E_x$ are shown in Fig. 3a for the glancing incidence. We choose three different points $A_1$, $A_2$ and $A_3$ denoted in Fig. 3c as examples. The corresponding simulated field spectra are presented in Fig 3b, showing excellent agreements with the experimental results. The measured peak positions of the $S_2$ and $A$ modes are 9.45 and 8.70 GHz, respectively. The spatial field profiles for the $S_2$ and $A$ modes are shown in Fig. 3c and 3d, respectively, and the corresponding characteristics can be easily identified. The $S_2$ mode at 9.45 GHz manifests the expected characteristics that the central antenna and side antennas are out of phase, while the $A$ mode at 8.70 GHz displays an anti-symmetric pattern. The spatial field profiles on the dashed lines indicated in Fig. 3c and 3d are plotted, respectively, in Fig. 3e and 3f. The experimental data represented by green dotted curves agree excellently with the simulated red solid curves.

**Dynamic eigenmode analysis of the system of antennas**. In this section, the theory of the dynamic eigenmode analysis for the system of antennas is outlined, and this method is similar to that for the system of plasmonic particles[41,42]. The corresponding quasi-static version is extensively used in searching certain eigenmodes, especially for split-ring structures[40]. Considering $N$ local current sources, $J_m^{(i)}$ located at $r_i'$ with $i=1, \ldots, N$ and $m=1, \ldots, M$ (index for the $m$-th harmonics), they are illuminated by an external field $\boldsymbol{E}^{\text{ext}}(\boldsymbol{r})e^{-i\omega t}$ (the factor $e^{-i\omega t}$ will be omitted below). The local current source $J_m^{(i)}$ is driven not only by the external field $\boldsymbol{E}^{\text{ext}}(r_i')$, but also by the fields radiated from all the sources $J_n^{(j)}$, including that from the others with $j \neq i$, and that from itself with $j=i$ as the antennas have a finite size. In the Methods, the Green function $G_{mn}(\boldsymbol{r}_i, \boldsymbol{r}_j, L_i, L_j)$ for the local current source is given in detail, where $L_i$ is the length of the $i$-th antenna. By applying this Green function to the Ohm's law, i.e., $\boldsymbol{J}=\sigma\boldsymbol{E}$, the equation of linear response can be expressed as $J_m^{(i)} = \sigma[E_{i,m}^{\text{ext}} + \sum_n \sum_j G_{mn}(r_i', r_j', L_i, L_j)J_n^{(j)}]$. Since we consider an open system here, the eigenstate with a purely real frequency $\omega$ does not exist in general. This equation can be rewritten in a matrix form as $\boldsymbol{MJ}=\boldsymbol{E}^{\text{ext}}$, or $\boldsymbol{J}=\boldsymbol{M}^{-1}\boldsymbol{E}^{\text{ext}}$ where $\boldsymbol{J}=(\cdots J_M^{(i-1)}, J_1^{(i)}, \cdots J_M^{(i)}, J_1^{(i+1)}, \cdots)^T$ and $\boldsymbol{E}^{\text{ext}}=(\cdots, E_M^{(i-1)}, E_1^{(i)}, \cdots, E_M^{(i)}, E_1^{(i+1)}, \cdots)^T$. The eigenvalues of $\boldsymbol{M}^{-1}$ (the inverse of the matrix $\boldsymbol{M}$) as a function of the frequency $\omega$ do give us the information of the spectral response of this system. These eigenvalues, denoted by $\sigma_\mu$ hereinafter, act as the mode conductivity for the corresponding eigenstates. The mode conductivity will exhibit a Lorentzian shape as a resonance takes place. The induced current $\boldsymbol{J}$ calculated from $\boldsymbol{M}^{-1}\boldsymbol{E}^{\text{ext}}$ can be further decomposed by using the eigenvectors $\left|\boldsymbol{J}^{(\mu)}\right\rangle$ corresponding to the eigenvalues $\sigma_\mu$. The coefficients $\left\langle \boldsymbol{J}^{(\mu)} \middle| \boldsymbol{E}^{\text{ext}} \right\rangle$ are the projection of the incident field onto the $\mu$-th eigenstate, representing the excitation efficiency of the external waves. This coefficient is obviously zero for an anti-symmetric mode at the normal incidence, such as the aforementioned $A$ mode.

The mode conductivities $\sigma_\mu$ for the three-antenna system are plotted as functions of the frequency in Fig. 4a. In our case, the lowest current harmonics dominates and we only keep the corresponding term $J_1^{(i)}$. Three peaks with different quality factors can be observed in Fig. 4a. By inspecting the symmetry of the corresponding eigenstates, these three peaks can be assigned to anti-symmetric $A$ mode (blue), symmetric $S_1$ (black) and

$S_2$ (red) modes. In Fig. 4a, the symmetry properties are indicated by the arrows pointing along the dipoles. The peak positions for the $A$, $S_1$ and $S_2$ modes are respectively 8.9, 10.2 and 9.64 GHz, which agree well with the peaks shown in the scattering efficiencies in Fig. 2a,c. We should emphasize that the mode conductivities $\sigma_\mu$ are intrinsic parameters of this system and do not depend on the external excitation. On the contrary, the projection coefficients $\langle J^{(\mu)} | E^{\text{ext}} \rangle$ are dependent on the external excitation. In Fig. 4b, we show the projection coefficients of the three different eigenstates for the glancing incidence. There is still a peak for the $A$ mode at about 8.9 GHz and a sharp peak for the $S_2$ mode at around 9.64 GHz. We note that there is also a small peak for the $S_1$ mode (electric dipole) accompanying with the toroidal dipole mode at 9.64 GHz. This comes from the fact that the total instantaneous current is not zero and it can give rise to a weak electric dipole. This accompanying electric dipole moment can also be observed in the radiation spectrum in Fig. 2b. The eigenmode analysis thus gives us a promising physics interpretation of the simulated and experimental results. We also note that the projection coefficient for $S_1$ mode does not have a peak at about 10.2 GHz corresponding to that in Fig. 4a, implying that the electric dipole mode cannot be excited efficiently at the glancing incidence. This is also consistent with the simulated results shown in Fig. 2a.

**From dipole antennas to metal nanoparticles**. Our theory can be generalized to a system of metal nanoparticles and thereby realizing toroidal modes at the optical frequencies with the same model. In this system, it is possible to access the deep subwavelength scale since the electric dipoles can be easily achieved in the nanoscale. With electric dipoles $p^{(i)}$ replacing the role of local current sources $J_m^{(i)}$, the theoretical analysis of this system parallels the dynamic eigenmode analysis for the system of antennas. Since the size of plasmonic particles is much smaller than the incident wavelength, we only consider the lowest angular momentum channel and the subscript $m$ is omitted. We choose the Drude-type permittivity for the metal: $\varepsilon(\omega) = 1 - \omega_p^2/(\omega^2 + i\gamma\omega)$, where $\omega_p$ is the plasma frequency and $\gamma$ is the electron scattering rate. The polarizability of $\alpha_i(\omega)$ for the $i$-th particle can be associated with the electric dipole term of the Mie coefficient $a_1$ by $\alpha_i(\omega) = (3i/2k_0^3) a_1^{(i)}(\omega)$, where $k_0 = \omega/c$ is the wave vector in vacuum, $c$ is the speed of light. Illuminated by the external field $E^{\text{ext}}$, the dipole moment $p^{(i)}$ can be determined by the coupled dipole equation[41,42]: $p^{(i)} = \alpha_i(\omega)[\sum_{j \neq i} g(r_i' - r_j') p^{(j)} + E_i^{\text{ext}}]$, where $g(r_i - r_j)$ is the dyadic Green function. By defining the following vectors $P = (\cdots, p_{i-1}, p_i, p_{i+1}, \cdots)^T$ and $E^{\text{ext}} = (\cdots, E_{i-1}, E_i, E_{i+1}, \cdots)^T$, we can again reformulate the coupled dipole equation to a compact one: $MP = E^{\text{ext}}$, or $P = M^{-1} E^{\text{ext}}$. The eigenvalue of $M^{-1}$, denoted by $\alpha_\mu$, can be treated as the mode polarizability of the $\mu$-th eigenmode for this system.

The first plasmonic system we considered is illustrated in the inset of Fig. 5a, similar to the structure shown in Fig. 1b. The central spherical particle has a radius different from that of the side ones in order to provide the "outflowing" current in Fig. 1b. We choose $r_1 = r_3 = 0.1$, $r_2 = 0.11$ ($2\pi c/\omega_p$) here, the unit $2\pi c/\omega_p$ is about 200 nm for $\omega_p = 6.18$ eV[41]. The center-to-center distance between the neighboring particles is $d = 0.3$ ($2\pi c/\omega_p$), and the dissipation is neglected here. The scattering efficiency for the glancing incidence with the incident wave $E_{\text{in}} = \hat{e}_z E_0 e^{ik_0 y}$ is shown in Fig. 5a. The sharp peak with frequency $\omega \sim 0.533 \omega_p$ corresponds to the $S_2$ mode discussed previously. The radiation spectrum for this mode possesses a peak for the toroidal dipole moment as shown by the red curve in Fig. 5b. A broader peak can be observed as $\omega \sim 0.543 \omega_p$ in Fig. 5a, which corresponds to the anti-symmetric mode ($A$ mode). The radiation spectrum has a peak of magnetic dipole resonance for this $A$ mode, shown by the blue curve in Fig. 5b. A tiny peak corresponding to the ordinary $S_1$ mode can be observed as $\omega \sim 0.562 \omega_p$ in the scattering spectrum.

The scattering and radiation spectra can be again interpreted by the eigenmode analysis. The imaginary parts of the eigenvalues of the matrix $M^{-1}$, Im($\alpha_\mu$), are plotted in Fig. 5c. The resonant peaks corresponding to that in Fig. 5a can be observed clearly at 0.536, 0.546 and 0.564$\omega_p$, for the $S_2$, $A$ and $S_1$ modes, respectively. The resonant $S_2$ mode is labelled as $S_{2R}$ mode, which has the maximum Im($\alpha_{s2}$). For the $S_{2R}$ mode, we assume that the corresponding eigenvector of the equation $MP = \alpha_{2R} P$ is $P = (p_1, p_2, p_3)$. The currents induced from these dipoles are simply: $j_i = \partial p_i / \partial t$. When the outflowing current from the central particle is equal to the inflowing currents from the side particles, namely, $p_2=-(p_1+p_3)$, this mode becomes a pure toroidal dipole mode as mentioned previously. This condition is equivalent to the vanishing of the total electric dipole moment, i.e., $P_{tot}=0$. In the quasi-static limit, this condition can be satisfied rigorously, which is $r_2=(15/8)^{1/3}r_1$. In the real situation with the retardation effects taken into account, $P_{tot}$ cannot be exactly zero. However, by altering the geometric parameters, optimization can be done for the minimum of $|P_{tot}|$. In Fig. 5d, $|P_{tot}|$ as a function of the radius $r_2$ is plotted. The minimum value of $|P_{tot}|$ is indicated by the dotted line. The corresponding $\text{Im}(\alpha)$ for the $S_{2R}$ mode is also shown. The minimum of $|P_{tot}|$ occurs near the maximum of $\text{Im}(\alpha)$, which implies that the response of $S_{2R}$ mode is still strong enough even if the total electric dipole moment $|P_{tot}|$ is close to zero.

The toroidal dipole resonance persists in the plasmonic system if the dissipation is taken into account as shown by the second example in Fig. 5e,f. In addition, we also replace the spherical particles by the cylindrical rods since they are more feasible in the standard fabrication. Here, the electron scattering rate is assumed to be $\gamma=0.003\omega_p$. The radius of the three rods are set to be the same, $r_1=r_2=r_3=0.1$ $(2\pi c/\omega_p)$, and the heights for the side rods are $L_1=L_3=0.2$ $(2\pi c/\omega_p)$. The height of central rod is chosen to be 0.21 $(2\pi c/\omega_p)$. At the frequency $\omega\sim0.442\omega_p$, a resonant toroidal mode can be observed clearly.

**Anapoles.** Another point worth noting is the interference between the electric dipole $P$ and toroidal dipole $T$ [34]. In the far field, the total scattered field from $P$ and $T$ can be written as $E_{sca}\sim e_r\times P\times e_r+ik_0 e_r\times T\times e_r$, where $e_r$ is the unit vector in the radial direction. Radiation fields from $P$ and $T$ will have constructive (destructive) interference when $P$ and $T$ oscillate with a phase difference $\pi/2$ ($-\pi/2$). When the radiation fields from $P$ and $T$ cancel each other, a radiation dip will occur, manifesting an "anapole" with strong field localization near the scatterers and no radiation leaks out.

Such "anapoles" can be found in both systems studied here. For the three-antenna system, we choose the length of the central antenna as $L_2=12$ mm and keep the other parameters the same as that in Fig. 2. In Fig. 6a, the result of the total radiation power of both $P$ and $T$ is shown by the red curve. The radiation power drops to zero at $f=10.4$ GHz as marked by an arrow, indicating the "anapole" phenomenon. The individual radiation power of $P$ and $T$ are plotted by the black and blue dashed curves, respectively. We note that these two curves also intersect at $f=10.4$ GHz, indicating that the radiated fields from $P$ and $T$ are the same in amplitude but opposite in sign. The corresponding normal component of the magnetic field $H_x$ of the anapole is shown in the inset, where the strong localization of the field is also seen. In Fig. 6b, we show the similar results for the three-metal-nanoparticle system. The radius of the central spherical particle is chosen to be $r_2=0.16\,(2\pi c/\omega_p)$, and the center-to-center distance between the adjacent spheres is $d=0.4\,(2\pi c/\omega_p)$. Other parameters are kept the same as that in Fig. 5a. An anapole is clearly seen at $\omega=0.543\,\omega_p$.

## DISCUSSIONS

The radiation field of the toroidal dipole in the far field gives the same pattern as that of the electric dipole, and so if one is only interested in the far-field radiation pattern, the notion of toroidal moment is probably not a necessity. However, if we consider the local current distribution of the object giving rise to the radiation field, such moment can be treated as a higher multipole coming from the contraction of the third rank tensor as we expand the local current density[6]. This is also the reason why the moments $P$ and $ik_0 T$ play the same role in the far-field radiation. In the quasi-static limit $k_0\rightarrow 0$, the toroidal dipole moment serves as a high order correction of the electric dipole moment. It will typically become more significant as the size of the scatterer increases, especially as the size is comparable to the wavelength[34]. In metamaterials, the resonating microstructures are usually subwavelength so that effective medium concepts can be applied and as such, toroidal moments are typically very small at the optical frequencies. In our proposal, the toroidal resonant mode can be achieved even if the incident wavelength is 3 times larger than the size of the structure in the example shown in Fig. 5a and 5e. The size of the nanoparticles can be further scaled down and the resonant peak still persists near the dipole resonance of the single particle.

In summary, we show that toroidal dipole resonances can exist in a minimal discrete model, consisting of three aligned electric dipoles with inversion symmetry. Such minimal model supports toroidal resonances from microwave up to optical frequencies.

This model is experimentally realized by three λ/2 antennas in the microwave regime. The measured near field spectra agree very well with the simulated results. The toroidal and magnetic dipole resonances can be directly observed from the measured spatial field profiles. The eigenmodes of this system are classified by symmetry and further interpreted by the dynamic eigenmode analysis analogous to the coupled dipole equations. This model can be generalized to other frequency regimes and two three-metal-nanoparticle systems have been investigated as examples. The realization of the toroidal dipole resonances in such simple systems may render possible applications in the control of light emission, optical sensing, and photoluminescence.

## METHODS

**Linear response of a wire antenna in the external field.** We consider a cylindrical antenna with a length $L$ and radius $\rho_0$, lying in the region $z \in [-L/2, L/2]$. The current oscillating along the antenna is assumed to be: $\boldsymbol{J}(z) = \hat{e}_z J(z) \pi \rho_0^2 \delta(x) \delta(y)$, where the current is treated as a line current for a distant observer. It is obvious that $J(z)$ should be zero at the end of the antenna. As a result, $J(z)$ can be expanded by the Fourier harmonics: $J(z) = \sum_m J_m \cos(zm\pi/L)$, where $m$ is an odd positive integer. For an expansion by a complete basis, the harmonics $J_n \sin(2zn\pi/L)$ with odd symmetry should be also incorporated. However, numerical calculations show that the contributions from these harmonics are extremely small under the normal incidence of plane waves. Therefore we neglect the terms of $J_n \sin(2zn\pi/L)$. The amplitude of $m$-th harmonics, $J_m$, is given by:

$$J_m = \frac{2}{L} \int_{-L/2}^{L/2} J(z) \cos(zm\pi/L) dz. \qquad (1)$$

In the cylindrical coordinates ($\rho$, $\phi$, $z$), the vector potential $\boldsymbol{A}$ induced by the $m$-th current harmonics is:

$$\boldsymbol{A}_m(\rho, \phi, z) = \hat{e}_z \frac{\mu_0 \rho_0^2}{4} \int_{-L/2}^{L/2} J_m \cos(z'm\pi/L) e^{ik_0 \sqrt{\rho^2 + (z-z')^2}} / \sqrt{\rho^2 + (z-z')^2} \, dz', \qquad (2)$$

which is independent on the azimuth angle $\phi$. The $z$ component of the electric field can be found through $\boldsymbol{A}$:

$$E_{z,m} = -\frac{iZ_0}{k_0}\frac{1}{\mu_0}\frac{1}{\rho}\frac{\partial}{\partial\rho}\left(\rho\frac{\partial A_{z,m}}{\partial\rho}\right), \quad (3)$$

where $Z_0 = \sqrt{\mu_0/\varepsilon_0}$ is the impedance of the free space. If we consider only a single antenna, the radiated electric field will react on itself. However, the electric field will diverge at the origin. As a reasonable approximation, we assume that it is the electric field $E_z$ on the boundary of the antenna that acts on the current source itself. The electric field at $\rho=\rho_0$ can be expanded by the Fourier series again:

$$E_{z,m}(\rho_0) = \sum_n E_{z,nm}\cos(zn\pi/L), \quad (4)$$

which implies that the source, i.e., the $m$-th current harmonics $J_m$, can induce many field harmonics indexed by $n$. The Fourier coefficients are given by:

$$E_{z,nm} = -\frac{iZ_0\rho_0^2}{4k_0}\frac{2}{L}\int_{-L/2}^{L/2}\left(\int_{-L/2}^{L/2} J_m \cos(z'm\pi/L) f^A(\rho_0,z,z')dz'\right)\cos(zn\pi/L)dz, \quad (5)$$

where

$$f^A(\rho,z,z') = \frac{1}{\rho}\frac{\partial}{\partial\rho}\left(\rho\frac{\partial}{\partial\rho}\left(\frac{e^{ik\sqrt{\rho^2+(z-z')^2}}}{\sqrt{\rho^2+(z-z')^2}}\right)\right). \quad (6)$$

The current and electric field can be related by the Ohm's law. The electric field consists of two parts: the external one and the one induced by the current itself, namely:

$$J_n \cos(zn\pi/L) = \sigma\left(\sum_m E_{z,nm}\cos(zn\pi/L) + E_n^{\text{ext}}\cos(zn\pi/L)\right). \quad (7)$$

The above equation can be written in a compact form by introducing a Green function:

$$\sigma^{-1}J_n - \sum_m G_{nm}(\rho_0,L,L)J_m = E_n^{\text{ext}}, \quad (8)$$

where the Green function is given as follows:

$$G_{nm}(\rho_0,L,L) = -\frac{iZ_0}{2\pi k_0 L}\int_{-L/2}^{L/2} F_m(\rho_0,z,L)\cos(zn\pi/L)dz, \quad (9)$$

and $F_m(\rho_0,z,L) = \pi\rho_0^2\int_{-L/2}^{L/2}\cos(z'm\pi/L)f^A(\rho_0,z,z')dz'$.

Eq. (8) can be further generalized to the system of multiple antennas:

$$\sigma^{-1}J_m^{(i)} - \sum_n\sum_j G_{mn}(r_{ij},L_i,L_j)J_n^{(j)} = E_{i,m}^{\text{ext}}, \quad (10)$$

where the subscripts *i* and *j* are the indices for the antennas, and the Green function has the following form:

$$G_{mn}(r_{ij}, L_i, L_j) = -\frac{iZ_0}{4\pi k_0} \frac{2}{L_i} \int_{-L_i/2}^{L_i/2} F_n(|r_i - r_j|, z, L_j) \cos(zm\pi/L_i) dz, \quad (11)$$

where |$r_i$-$r_j$| is the center-to-center distance between the antennas $A_i$ and $A_j$ as $i \neq j$, however, we choose $r_{ii} = \rho_0$ as $i = j$ (the "on-site energy").

**Radiation power of the multipole moments.** For a far-field observer, the field radiated by the local source can be ascribed to different multipoles. The multipole moments are defined by the current source, which are listed as follows[8]:

Electric dipole moment: $\boldsymbol{P} = -\frac{1}{i\omega}\int \boldsymbol{j} d^3 r$;

Magnetic dipole moment: $\boldsymbol{M} = \frac{1}{2c}\int (\boldsymbol{r} \times \boldsymbol{j}) d^3 r$;

Toriodal dipole moment: $\boldsymbol{T} = \frac{1}{10c}\int [(\boldsymbol{r} \cdot \boldsymbol{j})\boldsymbol{r} - 2r^2 \boldsymbol{j}] d^3 r$;

Electric quadrupole moment:

$$Q_{\alpha\beta} = -\frac{1}{2i\omega}\int \left[ r_\alpha j_\beta + r_\beta j_\alpha - \frac{2}{3}(\boldsymbol{r} \cdot \boldsymbol{j})\delta_{\alpha\beta} \right] d^3 r;$$

Magnetic quadrupole moment:

$$M_{\alpha\beta} = \frac{1}{3c}\int \left[ (\boldsymbol{r} \times \boldsymbol{j})_\alpha r_\beta + (\boldsymbol{r} \times \boldsymbol{j})_\beta r_\alpha \right] d^3 r.$$

The radiation power for the above multi-poles moments is given by:

$$I_p = \frac{2\omega^4}{3c^3}|\mathbf{P}|^2, \quad I_m = \frac{2\omega^4}{3c^3}|\mathbf{M}|^2, \quad I_t = \frac{2\omega^6}{3c^5}|\mathbf{T}|^2,$$
$$I_{Q_e} = \frac{\omega^6}{5c^5}\sum|Q_{\alpha\beta}|^2, \quad I_{Q_M} = \frac{\omega^6}{40c^5}\sum|M_{\alpha\beta}|^2.$$

### ACKNOWLEDGEMENTS

This work is supported the National Natural Science Foundation of China (Grant No. 11574037) and the Fundamental Research Funds for the Central Universities (Grant No. CQDXWL-2014-Z005). Work in Hong Kong is supported by Hong Kong Research Grant Council under Grants No. AoE/P-02/12. We thank Prof. S. T. Chui, Drs. Q. H. Liu and K. Ding for helpful discussions.

**FIGURES:**

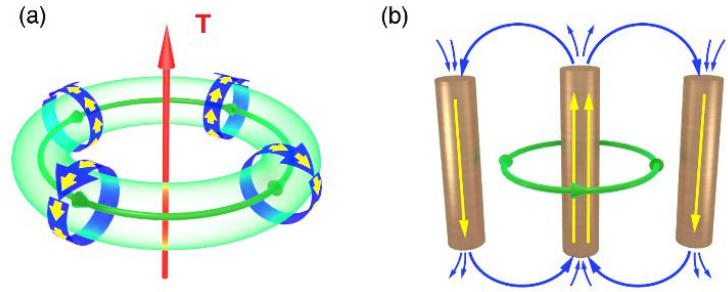

**Figure 1 | Toroidal moment from a discrete model based on resonant electric dipoles.** (a) Schematic view of a typical toroidal dipole. The currents are circulating along the meridians of the torus, indicated by the blue arrow. The induced magnetic dipoles are represented by the green arrows. The currents (blue arrows) can be replaced by discretized "current elements" shown by the yellow arrows. The *minimal* version of this discretization is given in (b), in which three electric dipoles serve as the "current elements". The alternating current outflows from the central dipole and inflows back from the side ones.

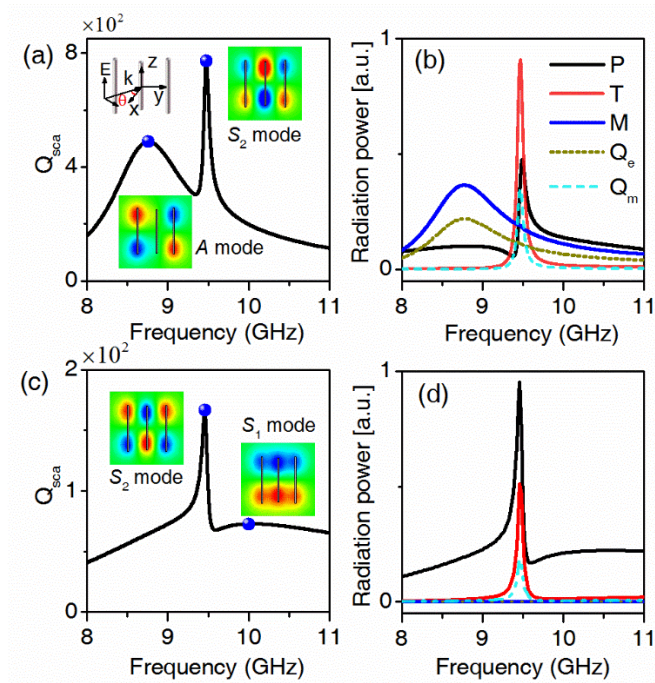

**Figure 2 | Toroidal mode in the three-antenna system.** The three-antenna system is sketched in the upper left inset in (a). The antennas pointing along the $z$ axis are aligned on the $y$ axis with inversion symmetry. An EM wave $\boldsymbol{E}_{in}=\hat{\boldsymbol{e}}_z E_0 e^{i(k_x x+k_y y)}$ is incident with an angle $\theta$ to the $x$ axis. (a) and (b) are for the glancing incidence with $\theta=\pi/2$. The simulated scattering efficiency is shown by the black curve in (a). The spatial field profiles for the two peaks marked by the blue dots are shown in the corresponding insets. The radiation spectra for different multipole moments: $\boldsymbol{P}$, $\boldsymbol{T}$, $\boldsymbol{M}$, $\boldsymbol{Q}_e$, $\boldsymbol{Q}_m$, are plotted in (b). The

scattering peaks at 8.75 and 9.48 GHz in (a) correspond to the resonant excitation of the magnetic and toroidal dipole modes, respectively. Similar simulated results are given in (c) and (d) for the normal incidence with $\theta=0$. The electric dipole resonance at about 10.0 GHz can be observed as $\theta=0$. The simulations are conducted by COMSOL multi-physics.

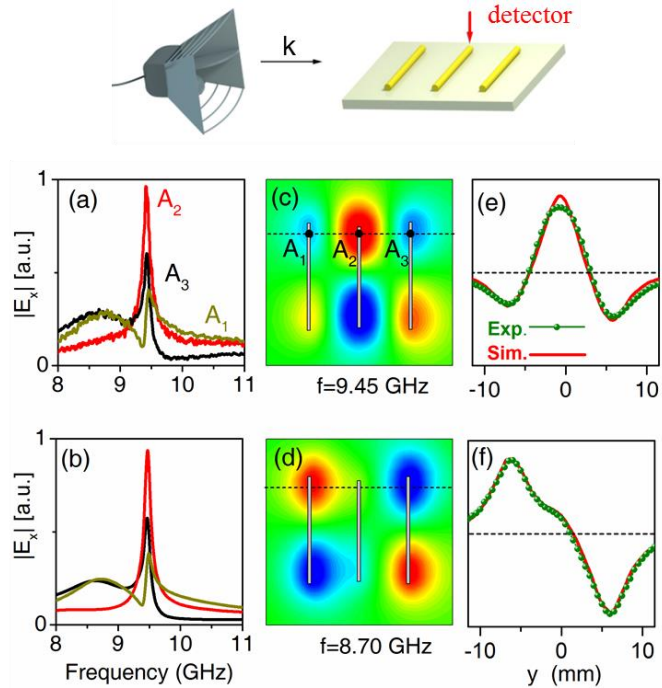

**Figure 3 | Measurements of the properties of the toroidal and magnetic dipole modes.** The experimental setup is illustrated in the upper inset. The measured and simulated near field spectra for the glancing incidence are shown in (a) and (b), respectively. The yellow, red and black curves are measured, respectively, at the points $A_1$, $A_2$ and $A_3$ indicated in (c). The measured spatial field profile for the $S_2$ mode at the frequency $f=9.45$ GHz is shown in (c), while that for the $A$ mode at $f=8.70$ GHz is shown in (d). The spatial field profiles along the black dashed lines displayed in (c) and (d) are plotted by the green dotted curves in (e) and (f), respectively. The corresponding simulated results are given by the red solid curves. It can be observed clearly that the central antenna is out of phase with the side ones in (c) and (e) for the toroidal dipole mode, while the field pattern is anti-symmetric in (d) and (f) for the magnetic dipole mode.

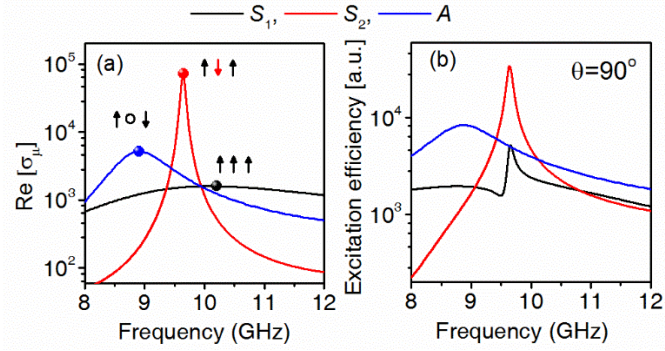

**Figure 4 | Dynamic eigenmode analysis for the three-antenna system.** The real parts of the mode conductivities $\sigma_\mu$ for the three eigenmodes are plotted in (a). The relative phases of the dipoles are shown by the arrows. The peaks of the anti-symmetric mode (*A* mode), symmetric modes ($S_1$ and $S_2$) located at, respectively, $f$=8.9, 10.2 and 9.64 GHz, agree well with the peaks of the scattering spectra shown in Fig. 2. The projection coefficients of the three modes for the glancing incidence are shown in (b). The large peak of the projection coefficient for the $S_2$ mode demonstrates that the coupling between the $S_2$ mode and the incident wave with $\theta=\pi/2$ is strong, giving rise to the resonant excitation of the toroidal mode.

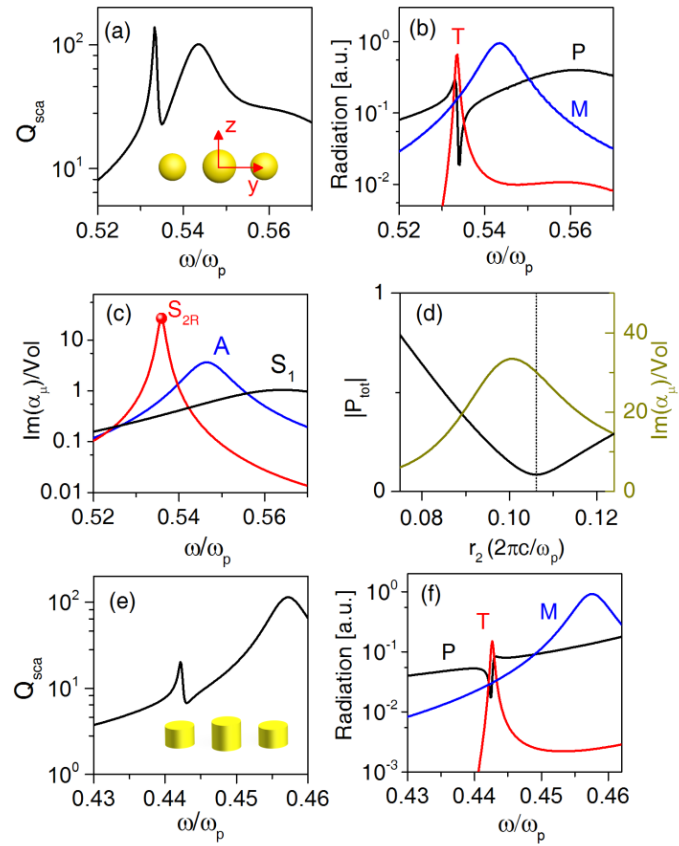

**Figure 5 | Toroidal modes in the system of metal nanoparticles.** For the system of three plasmonic spheres, the scattering spectrum is plotted by the black curve in (a). Two

peaks can be observed, i.e., the one at $\omega=0.533\omega_p$ and the one at $\omega=0.543\omega_p$. The radiation from toroidal and magnetic dipoles that give rise to these two peaks are shown by the red and blue curves in (b), respectively. The imaginary parts of the mode polarizability $\alpha_\mu$ for the three eigenmodes are given in (c). The $S_{2R}$ state, marked by the red dot, is the resonant $S_2$ mode. The total electric dipole moment $|\bm{P}_{tot}|$ for $S_{2R}$ state is plotted in (d) as we vary the radius $r_2$ of the central particle. The minimum of $|\bm{P}_{tot}|$ corresponds to the purest toroidal dipole. In (e) and (f), another example composed of three cylindrical rods is also given, in which the dissipation of the metal has been taken into account with $\gamma=0.003\omega_p$.

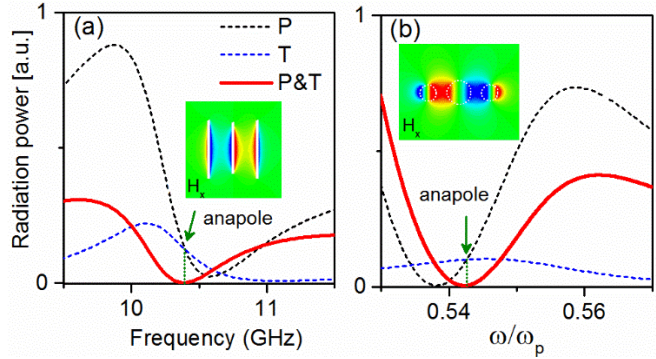

**Figure 6 | "Anapoles" in both the systems of dipole antennas and plasmonic particles.** The radiation spectra are shown in (a) and (b), respectively. The length of the central antenna is chosen to be $L_2=12$ mm in (a), and other parameters are kept the same as those in Fig. 2. The radius of the central sphere is set to be $r_2=0.16\,(2\pi c/\omega_p)$, and the center-to-center distance between the adjacent spheres is $d=0.4\,(2\pi c/\omega_p)$. Other parameters are kept the same with those in Fig. 5a.